\documentclass[fleqn,10pt]{wlscirep}
\usepackage[utf8]{inputenc}
\usepackage[T1]{fontenc}
\title{Agent-based model using GPS analysis for infection spread and inhibition mechanism of SARS-CoV-2 in Tokyo}

\author[1]{Taishu Murakami}
\author[1*]{Shunsuke Sakuragi}
\author[2]{Hiroshi Deguchi}
\author[1]{Masaru Nakata}
\affil[1]{MRI Research Associates, Inc., 2-10-3 Nagata-cho, Chiyoda-ku, Tokyo 100-0014 Japan.}
\affil[2]{Faculty of Commerce and Economics, Chiba University of Commerce, 1-3-1 Konodai, Ichikawa-shi, Chiba 272-8512 Japan.}

\affil[*]{shunsuke\_sakuragi@mri-ra.co.jp}

\begin{abstract}
Analyzing the SARS-CoV-2 pandemic outbreak based on actual data while reflecting the characteristics of the real city provides beneficial information for taking reasonable infection control measures in the future. We demonstrate agent-based modeling for Tokyo based on GPS information and official national statistics and perform a spatiotemporal analysis of the infection situation in Tokyo. As a result of the simulation during the first wave of SARS-CoV-2 in Tokyo using real GPS data, the infection occurred in the service industry, such as restaurants, in the city center, and then the infected people brought back the virus to the residential area; the infection spread in each area in Tokyo. This phenomenon clarifies that the spread of infection can be curbed by suppressing going out or strengthening infection prevention measures in service facilities. It was shown that pandemic measures in Tokyo could be achieved not only by strong control, such as the lockdown of cities, but also by thorough infection prevention measures in service facilities, which explains the curb phenomena in real Tokyo.

\end{abstract}

\begin{document}

\flushbottom
\maketitle

\thispagestyle{empty}

\section*{Introduction}

In Japan, to control the SARS-CoV-2 pandemic that occurred in the spring of 2020, the government actively isolated infected people. Additionally, workers in the service industry were requested to take leaves of absence, shorten working hours, and refrain from going out. Although the spread of the infection occurred, it was successfully controlled. The number of newly infected people per day was similar to that of New Zealand, which implemented a lockdown\cite{Nishiura_NZ}, at less than 50 infections per day.
The Japanese government repeated these weak measures to cope with the pandemic while activating economic activities. As a result, there was an exponential increase in the number of infected people after July 2020. The number of infected people per day did not drop to the same level as when the first spikes occurred; however, these weak measures have effectively controlled the SARS-CoV-2 pandemic in Japan.

Because Japan was able to suppress infection with weak infection control measures compared to other countries, analyzing infection spread and control mechanisms in Japan provides epidemiological knowledge regarding infection control measures. Agent-based model simulation, which is an approach for advancing evidence-based medicine (EBM+), is suitable for analysis focusing on the mechanism of infection spread and suppression \cite{WilliamsonA13}. 
 In contrast to macro models such as the susceptible-exposed-infectious-recovered (SEIR) model, micromodels such as agent-based models can incorporate information on actual city structures (e.g., cohort structures, geographic information, and national statistical surveys)\cite{ferguson_2005,ferguson_2006,Germann5935} and people’s behavior (e.g., information from GPS data)\cite{wang_heterogeneous_2021}. Thus, it is possible to analyze the pandemic phenomena\cite{10.1371/journal.pone.0247182}.

In this study, to clarify the mechanism of SARS-CoV-2 infection spread and control in Japan, we performed an agent-based infection simulation using Tokyo GPS information, where the pandemic most notably occurred. The simulation results show that the general tendency for the infection was to spread from the city center to the suburbs, as illustrated by the infection in the Tokyo service industry (e.g., restaurants), where it was brought to the suburban residential area during the SARS-CoV-2 pandemic. The analysis of GPS information and simulation results showed that the suppression of human flow as part of infection control measures caused a decrease in the number of consumers in the service industry, resulting in controlling the infection. In addition, the scenario analysis revealed that strengthening infection prevention measures in the service industry, which has become a hub for the spread of infection, will significantly suppress infection. Our present findings indicate that, in the early stage of a pandemic, promoting the prevention of infection in individuals in the service industry, in addition to the weak suppression of traffic, has the same effect as a lockdown.

\section*{Results}
In analyzing pandemic mechanisms, it is difficult to identify the date and location of the infection in epidemiological studies. Our present simulation method, using GPS information with detailed human information, overcomes this problem. The simulation was performed for the first wave of the SARS-CoV-2 pandemic in Tokyo from March to May 2020. The number of initially infected persons was 15 to simulate the spread of infection. Weighting was performed according to the area of residence of the affected persons on March 16. The infected state was assigned to the agent as a random number. For our present method, we must fix the infectivity/filter parameter $k_u$ (see Methods). Thus, we fitted the time dependence of the total number of infected people using an epidemic curve that reflected the time-lag factor in the epidemiological survey. Figure 1 shows the time dependence of the number of symptomatic patients obtained from the simulation. The spread and convergence of the infection were observed owing to the agent’s behavior, which was time-modulated by GPS data. In addition, our present calculation reproduced the estimates of the onset date of infection estimated from the inspection date by Ref.\cite{Nishiura_Rt} by tuning $k_u$.

In addition, one of the merits of using GPS information is that it is possible to precisely determine places where people stay, such as entertainment districts and residential areas\cite{VYKLYUK2021103662}. Thus, the infection risk and number of local reproductions by time and facility can be calculated. By analyzing the increasing tendency of the number of infected people by region, a strong correlation with the actual data was shown. Therefore, it can be said that this simulation model reproduces the infection tendency of the first wave of SARS-CoV-2 in Tokyo.

To understand the risk of infection at each facility in Tokyo, we defined contact, which is the main risk factor, as "people approaching each other within a radius of 1 meter” and calculated this value per 1 hour, i.e., contact frequency for the daytime and nighttime in the third week of March. In addition, we defined the local effective reproduction number Rt as the maximum value at which one infected person generated a new infected person per hour at each facility in the cell. Figure 2 shows the calculation results for the contact frequency and $R_t$. Contact frequency and Rt tended to be high in the office district on weekdays and during the day. In addition, it was found that facilities with a contact frequency above the average value for Tokyo were part of the service industry in the downtown area. The results show that office workers’ use of facilities in densely populated areas increases the infection risk.

\section*{Discussion}
Figure 3 shows the time dependence of the ratio of infected facilities to the number of days since the initial infection occurred in the simulation. It can be seen that infections in the home and service industries occur in sequence in an oscillating manner. These results indicate a chained spread of infection, such as infections originating in the service industry by office workers who frequently move between regions and spread when taken home. This tendency is comparable to the spread mechanisms reported for other infectious diseases\cite{yang_topological_2020}.

To clarify the mechanism of infection convergence, the difference between the daytime/nighttime population during the first to second weeks of April 2020 and that of 2019 (at the same time before the SARS-CoC-2 pandemic) is shown in Figure 4. This value shows the effect of "weak infection control measures" by the government in Tokyo. Furthermore, Figure 4 shows that self-restraint by citizens was carried out mainly in the office district and the surrounding downtown area. This result indicates that the weak infection control measures, that of requests to refrain from going out and doing business with the service industry in Japan, led to less crowding and the suppression of commercial use, which is a hub for the spread of infections. Thus, the first wave converged according to the above measures.

To verify this result, the relationship between the infections that occurred between the agents, that is, the infection network, is represented by a directed graph (Figure 5). As shown in Figure 5, the number of infected people increased exponentially in the home and service industries in the early stage of the spread of infection, the rate of infection in the service industry decreased after refraining from activities, and the spread of infection converged. In other words, by refraining from going out, infections in the service industry were shut out. Thus, the infection in the service industry, which is the hub of infection on the go, was suppressed, which is essential for the convergence of infection in Tokyo.

It can be seen that the first wave of the pandemic in Japan was controlled immediately despite weak political intervention compared to other countries. Lockdown, which is a strong measure against SARS-CoV-2 infection, restricts the movement of citizens. Therefore, there are other factors that control the spread of infection in addition to the suppression of activity during the spread and convergence of the pandemic in Tokyo. In epidemiological studies and simulation analysis in Ref \cite{Utomo_2008, Deguchi_2014}, it has been proposed to classify the following four control modes to suppress the pandemic: (1) intervention control that suppresses the probability of infection, (2) intervention control that limits human behavior, (3) isolation control for infected persons, and (4) intervention control that raises antibody titers with vaccines. Furthermore, a survey showed that the rate of face mask wearing in public places in Japan was high even before the SARS-CoV-2 pandemic13; thus, intervention control that suppresses the probability of infection spread was already done naturally during the pandemic of SARS-CoV-2.

The simulation illustrates that the speed of the spread of infection was suppressed when the effect of the filter measures $k_u$ was enhanced, assuming that masks were worn in public places. In particular, it has become clear that the pandemic would not occur when the probability of infection is halved by assuming that masks were worn in the service industry, which is the hub of the above-mentioned infection. Therefore, thorough infection control measures in public places for individuals can have the same effect as a strong lockdown in cities, even with weak infection control measures such as self-restraint requests.

According to an analysis by the Japanese government\cite{mlit}, the request to refrain from going out during the first wave of the SARS-CoV-2 pandemic reduced the train utilization rate in urban areas by up to 70\% compared to pre-pandemic times. However, after the second wave, it decreased by approximately 40\% at the maximum compared to pre-pandemic times, and the degree of self-restraint of individuals by request gradually weakened. In Japan, the number of infected people has not reached the level of convergence of the first wave after the second wave of infection. Thus, this indicates that if an individual’s infection prevention measures and behavioral self-restraint reach a certain level, the effect is equivalent to that of a lockdown, and if either is lacking, a sufficient effect cannot be obtained.

In conclusion, we focused on infectious disease prevention measures, especially in service facilities (e.g., restaurants), among public facilities where contact between people occurs and pandemics, by controlling the local population to a certain extent and investigating the usage of agent-based model calculation. The results show that this measure significantly reduced the risk of infection. The results also highlight SARS-CoV-2 industry-specific measures, such as restricting restaurant operations and refraining from going out in the early stages of a pandemic have enough effect to avoid the spread of infection even if no lockdown is implemented. In addition, the number of infected people and the number of deaths were lower in Japan than those in other countries due to the suppression of infection, especially at facilities that became the hub of infection due to citizens’ self-action to lower the risk of infection. Our present analysis will be an important guideline for considering policies to be implemented in the event of a future pandemic.

It was also shown that the pandemic situation differs greatly depending on the city structure and the behavior of citizens. In addition, global pandemics, illustrated by the new influenza, occur once every 10 years. Therefore, by applying this simulation model to cities worldwide, appropriate policy decisions could be made promptly in the event of a pandemic, considering the urban structure at that time.

\section*{Methods}

To analyze the scenario of the new coronavirus infectious disease prevention measures in Tokyo, we built an artificial society that downscaled the daytime population in Tokyo to 100,000. In an artificial society, each agent has attributes such as age, sex, place of residence, and commuting/school place, which are generated by weighted random numbers based on national statistical data \cite{census}. The geographical information of Tokyo is represented by cells divided by a 500-meter mesh, and the establishments and residences are included in the major industrial classification based on the economic census activity survey (hereinafter, the establishments and residences on the cell are collectively grouped, and the industrial category and the residence to which the facility is applicable is called a facility type), and it is a cohort of urban functions. In addition, cells other than those in Tokyo were treated as the same cells for simplicity.

From the GPS data, we obtained the hourly dwellings, source cells, number of people, destination cells, and number of people in each cell. Based on this information, all agents were weighted by macroscopic statistical data on the time dependence of the employment rate in each industry and the going-out rate other than employment. For simplicity, we assume that all the agents are present in the residence cell at 3:00 am. Based on the macroscopic statistical data, the agents were set to start their homecoming behavior after their working hours and non-working out time and move to their residence after the time in the macroscopic statistical data of their homecoming time had elapsed. The behavioral patterns of the agents generated by this method are in agreement with the macroscopic statistical data of the employment rate and the out-of-home rate for each time and city, and it is confirmed that the agents adopt appropriate behavioral patterns.

The agent stochastically changes the infection state according to the infection mechanism described below. Figure 6 shows a transition diagram of the infection state of the agent.

$\phi$ was set to 0.875, and $z'$was set on the 4th based on the Japanese isolation rules at that time.
Additionally, $y_i$ and $z_i$ were generated from the following probability distributions based on previous reports \cite{donnelly_epidemiological_2003, ferguson2020report, doi:10.1126/science.abb5793}: 
\begin{equation}
y_i\backsim\mathrm{Gamm}\left(y;k=2.1,\theta=2\right)=\frac{y^{k-1}\mathrm{exp} \left(-\frac{y}{\theta}\right)}{\Gamma\left(\theta\right)\theta^k}, 
\end{equation}
\begin{equation}
z_i\backsim\mathrm{Pois}\left(z;\rho=7\right)=\frac{\rho^z\mathrm{exp} \left(-\rho\right)}{z!}.
\end{equation}

The agent infection probability $p_{mut}$ was introduced as a function of the amount of virus contamination $F_{mut}$ for the agent: 
\begin{equation}
p_{mut}=tanh\left(F_{mut}\right). 
\end{equation}

The amount of agent contamination $F_ {mut}$ at time t is the number of infectious agents $M_ {mut}$ in the facility (mesh code $m$, facility type $u$) in the cell and the total area of the facility $A_ {mu}$. The effect of infectious disease prevention measures (filter measures) at the facility was defined as follows using $k_u$:  
\begin{equation}
F_{mut}=k_u\frac{M_{mut}}{A_{mu}}.
\end{equation}

Here, it is assumed that the facility is sufficiently sterilized and ventilated and that the amount of contamination of the facility at time t decreased to a negligible amount after 1 hour. In addition, the agent pollution amount $F_ {mut}$ also treats the reduction rate of the pollution amount 1 hour ago as 1. The effect of filter measures $k_u$ includes the effects of disinfecting the hands and fingers of facility users and wearing masks. For simplicity, a uniform value was set for each facility type $u$.

Simulations were conducted multiple times, and it was confirmed that the trends of these simulations were qualitatively consistent. The average of five trials was used to calculate the number of infected individuals.

\bibliography{sample}

\begin{thebibliography}{10}
\urlstyle{rm}
\expandafter\ifx\csname url\endcsname\relax
  \def\url#1{\texttt{#1}}\fi
\expandafter\ifx\csname urlprefix\endcsname\relax\def\urlprefix{URL }\fi
\expandafter\ifx\csname doiprefix\endcsname\relax\def\doiprefix{DOI: }\fi
\providecommand{\bibinfo}[2]{#2}
\providecommand{\eprint}[2][]{\url{#2}}

\bibitem{Nishiura_NZ}
\bibinfo{author}{Kayano, T.} \& \bibinfo{author}{Nishiura, H.}
\newblock \bibinfo{journal}{\bibinfo{title}{A comparison of case fatality risk
  of covid-19 between singapore and japan}}.
\newblock {\emph{\JournalTitle{Journal of Clinical Medicine}}}
  \textbf{\bibinfo{volume}{9}}, \doiprefix\url{10.3390/jcm9103326}
  (\bibinfo{year}{2020}).

\bibitem{WilliamsonA13}
\bibinfo{author}{Williamson, J.}
\newblock \bibinfo{journal}{\bibinfo{title}{17 ebm+: increasing the systematic
  use of mechanistic evidence}}.
\newblock {\emph{\JournalTitle{BMJ Evidence-Based Medicine}}}
  \textbf{\bibinfo{volume}{24}}, \bibinfo{pages}{A13--A14},
  \doiprefix\url{10.1136/bmjebm-2019-EBMLive.25} (\bibinfo{year}{2019}).
\newblock \eprint{https://ebm.bmj.com/content/24/Suppl_1/A13.full.pdf}.

\bibitem{ferguson_2005}
\bibinfo{author}{Ferguson, N.~M.} \emph{et~al.}
\newblock \bibinfo{journal}{\bibinfo{title}{Strategies for containing an
  emerging influenza pandemic in southeast asia}}.
\newblock {\emph{\JournalTitle{Nature}}} \textbf{\bibinfo{volume}{437}},
  \bibinfo{pages}{209--214}, \doiprefix\url{10.1038/nature04017}
  (\bibinfo{year}{2005}).

\bibitem{ferguson_2006}
\bibinfo{author}{Ferguson, N.~M.} \emph{et~al.}
\newblock \bibinfo{journal}{\bibinfo{title}{Strategies for mitigating an
  influenza pandemic}}.
\newblock {\emph{\JournalTitle{Nature}}} \textbf{\bibinfo{volume}{442}},
  \bibinfo{pages}{448--452}, \doiprefix\url{10.1038/nature04795}
  (\bibinfo{year}{2006}).

\bibitem{Germann5935}
\bibinfo{author}{Germann, T.~C.}, \bibinfo{author}{Kadau, K.},
  \bibinfo{author}{Longini, I.~M.} \& \bibinfo{author}{Macken, C.~A.}
\newblock \bibinfo{journal}{\bibinfo{title}{Mitigation strategies for pandemic
  influenza in the united states}}.
\newblock {\emph{\JournalTitle{Proceedings of the National Academy of
  Sciences}}} \textbf{\bibinfo{volume}{103}}, \bibinfo{pages}{5935--5940},
  \doiprefix\url{10.1073/pnas.0601266103} (\bibinfo{year}{2006}).
\newblock \eprint{https://www.pnas.org/content/103/15/5935.full.pdf}.

\bibitem{wang_heterogeneous_2021}
\bibinfo{author}{Wang, H.}, \bibinfo{author}{Ghosh, A.}, \bibinfo{author}{Ding,
  J.}, \bibinfo{author}{Sarkar, R.} \& \bibinfo{author}{Gao, J.}
\newblock \bibinfo{journal}{\bibinfo{title}{Heterogeneous interventions reduce
  the spread of covid-19 in simulations on real mobility data}}.
\newblock {\emph{\JournalTitle{Scientific Reports}}}
  \textbf{\bibinfo{volume}{11}}, \bibinfo{pages}{7809},
  \doiprefix\url{10.1038/s41598-021-87034-z} (\bibinfo{year}{2021}).

\bibitem{10.1371/journal.pone.0247182}
\bibinfo{author}{Kirpich, A.} \emph{et~al.}
\newblock \bibinfo{journal}{\bibinfo{title}{Development of an interactive,
  agent-based local stochastic model of covid-19 transmission and evaluation of
  mitigation strategies illustrated for the state of massachusetts, usa}}.
\newblock {\emph{\JournalTitle{PLOS ONE}}} \textbf{\bibinfo{volume}{16}},
  \bibinfo{pages}{1--15}, \doiprefix\url{10.1371/journal.pone.0247182}
  (\bibinfo{year}{2021}).

\bibitem{Nishiura_Rt}
\bibinfo{author}{Jung, S.-m.}, \bibinfo{author}{Endo, A.},
  \bibinfo{author}{Kinoshita, R.} \& \bibinfo{author}{Nishiura, H.}
\newblock \bibinfo{journal}{\bibinfo{title}{Projecting a second wave of
  covid-19 in japan with variable interventions in high-risk settings}}.
\newblock {\emph{\JournalTitle{Royal Society Open Science}}}
  \textbf{\bibinfo{volume}{8}}, \bibinfo{pages}{202169},
  \doiprefix\url{10.1098/rsos.202169} (\bibinfo{year}{2021}).
\newblock
  \eprint{https://royalsocietypublishing.org/doi/pdf/10.1098/rsos.202169}.

\bibitem{VYKLYUK2021103662}
\bibinfo{author}{Vyklyuk, Y.}, \bibinfo{author}{Manylich, M.},
  \bibinfo{author}{Skoda, M.}, \bibinfo{author}{Radovanovic, M.~M.} \&
  \bibinfo{author}{Petrovic, M.~D.}
\newblock \bibinfo{journal}{\bibinfo{title}{Modeling and analysis of different
  scenarios for the spread of covid-19 by using the modified multi-agent
  systems – evidence from the selected countries}}.
\newblock {\emph{\JournalTitle{Results in Physics}}}
  \textbf{\bibinfo{volume}{20}}, \bibinfo{pages}{103662},
  \doiprefix\url{https://doi.org/10.1016/j.rinp.2020.103662}
  (\bibinfo{year}{2021}).

\bibitem{yang_topological_2020}
\bibinfo{author}{Yang, C.~H.} \& \bibinfo{author}{Jung, H.}
\newblock \bibinfo{journal}{\bibinfo{title}{Topological dynamics of the 2015
  south korea mers-cov spread-on-contact networks}}.
\newblock {\emph{\JournalTitle{Scientific Reports}}}
  \textbf{\bibinfo{volume}{10}}, \bibinfo{pages}{4327},
  \doiprefix\url{10.1038/s41598-020-61133-9} (\bibinfo{year}{2020}).

\bibitem{Utomo_2008}
\bibinfo{author}{Putro, U.~S.} \emph{et~al.}
\newblock \bibinfo{journal}{\bibinfo{title}{Searching for effective policies to
  prevent bird flu pandemic in bandung city using agent-based simulation}}.
\newblock {\emph{\JournalTitle{Systems Research and Behavioral Science}}}
  \textbf{\bibinfo{volume}{25}}, \bibinfo{pages}{663--673},
  \doiprefix\url{https://doi.org/10.1002/sres.948} (\bibinfo{year}{2008}).
\newblock \eprint{https://onlinelibrary.wiley.com/doi/pdf/10.1002/sres.948}.

\bibitem{Deguchi_2014}
\bibinfo{author}{Minh, D.} \emph{et~al.}
\newblock \bibinfo{journal}{\bibinfo{title}{An analysis on risk of
  influenza-like illness infection in a hospital using agent-based
  simulation}}.
\newblock {\emph{\JournalTitle{Public Health Frontier}}}
  \textbf{\bibinfo{volume}{14}}, \bibinfo{pages}{63--74}
  (\bibinfo{year}{2014}).

\bibitem{mlit}
\bibinfo{title}{{Ministry of Land, Infrastructure, Transport and Tourism}}.
\newblock
  \bibinfo{howpublished}{\url{https://www.mlit.go.jp/tetudo/tetudo_fr1_000062.html}}
  (\bibinfo{year}{{2021}}).

\bibitem{census}
\bibinfo{title}{{Economic and Financial Data for Japan}}.
\newblock
  \bibinfo{howpublished}{\url{https://www.soumu.go.jp/english/dgpp_ss/nsdp.htm}}
  (\bibinfo{year}{{2018}}).

\bibitem{donnelly_epidemiological_2003}
\bibinfo{author}{Donnelly, C.~A.} \emph{et~al.}
\newblock \bibinfo{journal}{\bibinfo{title}{Epidemiological determinants of
  spread of causal agent of severe acute respiratory syndrome in hong kong}}.
\newblock {\emph{\JournalTitle{The Lancet}}} \textbf{\bibinfo{volume}{361}},
  \bibinfo{pages}{1761--1766}, \doiprefix\url{10.1016/S0140-6736(03)13410-1}
  (\bibinfo{year}{2003}).
\newblock \bibinfo{note}{Publisher: Elsevier}.

\bibitem{ferguson2020report}
\bibinfo{author}{Ferguson, N.} \emph{et~al.}
\newblock \bibinfo{journal}{\bibinfo{title}{Report 9: Impact of
  non-pharmaceutical interventions (npis) to reduce covid19 mortality and
  healthcare demand}}.
\newblock {\emph{\JournalTitle{Imperial College London}}}
  \textbf{\bibinfo{volume}{10}}, \bibinfo{pages}{491--497}
  (\bibinfo{year}{2020}).

\bibitem{doi:10.1126/science.abb5793}
\bibinfo{author}{Kissler, S.~M.}, \bibinfo{author}{Tedijanto, C.},
  \bibinfo{author}{Goldstein, E.}, \bibinfo{author}{Grad, Y.~H.} \&
  \bibinfo{author}{Lipsitch, M.}
\newblock \bibinfo{journal}{\bibinfo{title}{Projecting the transmission
  dynamics of sars-cov-2 through the postpandemic period}}.
\newblock {\emph{\JournalTitle{Science}}} \textbf{\bibinfo{volume}{368}},
  \bibinfo{pages}{860--868}, \doiprefix\url{10.1126/science.abb5793}
  (\bibinfo{year}{2020}).
\newblock \eprint{https://www.science.org/doi/pdf/10.1126/science.abb5793}.

\end{thebibliography}

\section*{Acknowledgements}
We thank to Mitsubishi Research Institute, Inc. discussing the interpretation of the results of this study. 
We also thank to Location Mind Co., Ltd. in providing GPS data. 

\section*{Author contributions statement}
T.M. and S.S conceived the calculations, M.N., H.D and S.S. conducted the calculations, T.M. and S.S. analysed the results. T.M, S.S and H.D. wrote the manuscript. 
All authors reviewed the manuscript.

\begin{figure}[ht]
\centering
\includegraphics[width=\linewidth]{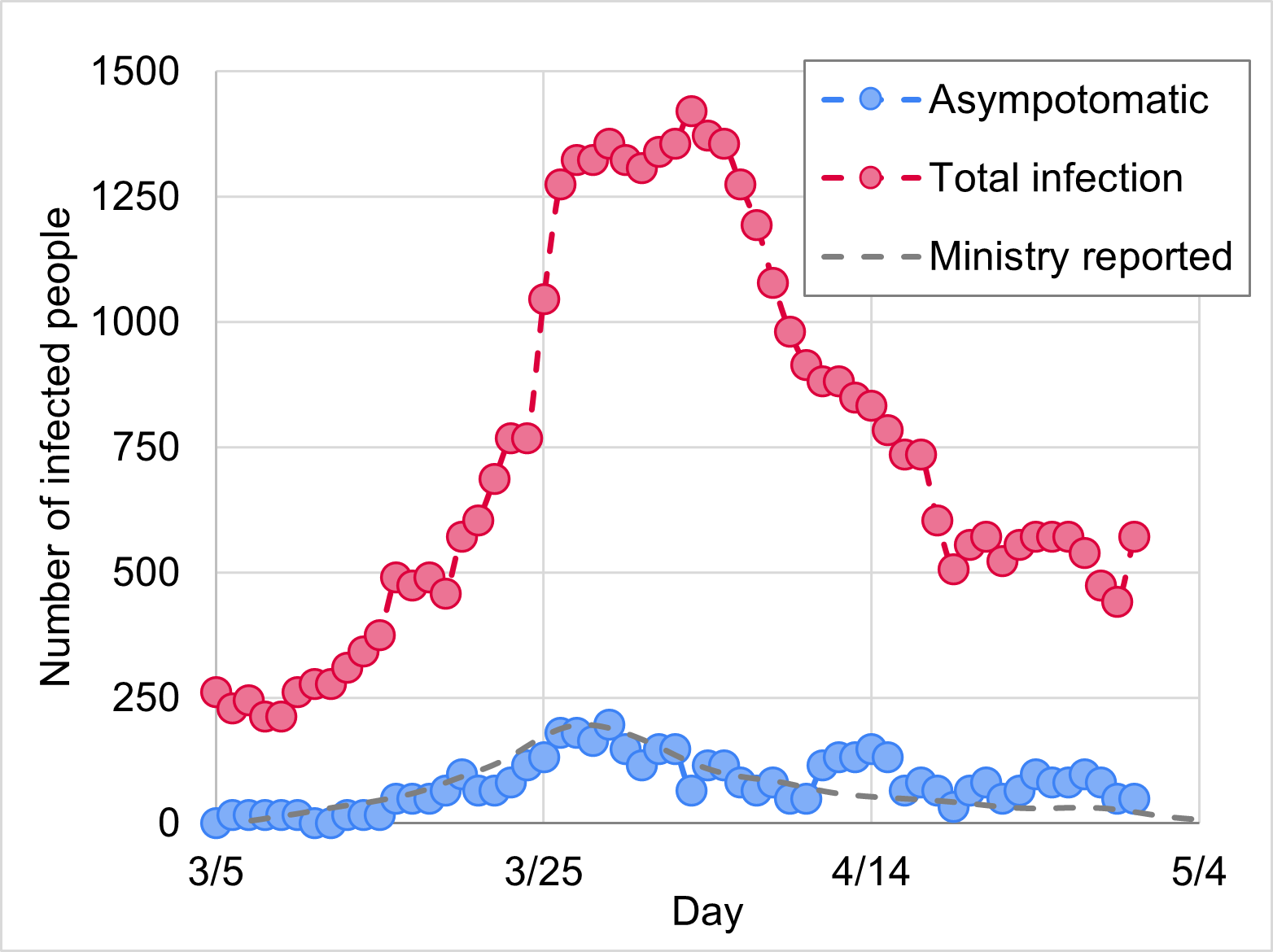}
\caption{
Epidemic curves obtained by calculation and values reported by the Japanese Ministry of Health, Labour, and Welfare. Ministry reported values are taken from Ref. 8.
}
\label{fig:stream}
\end{figure}

\begin{figure}[ht]
\centering
\includegraphics[width=\linewidth]{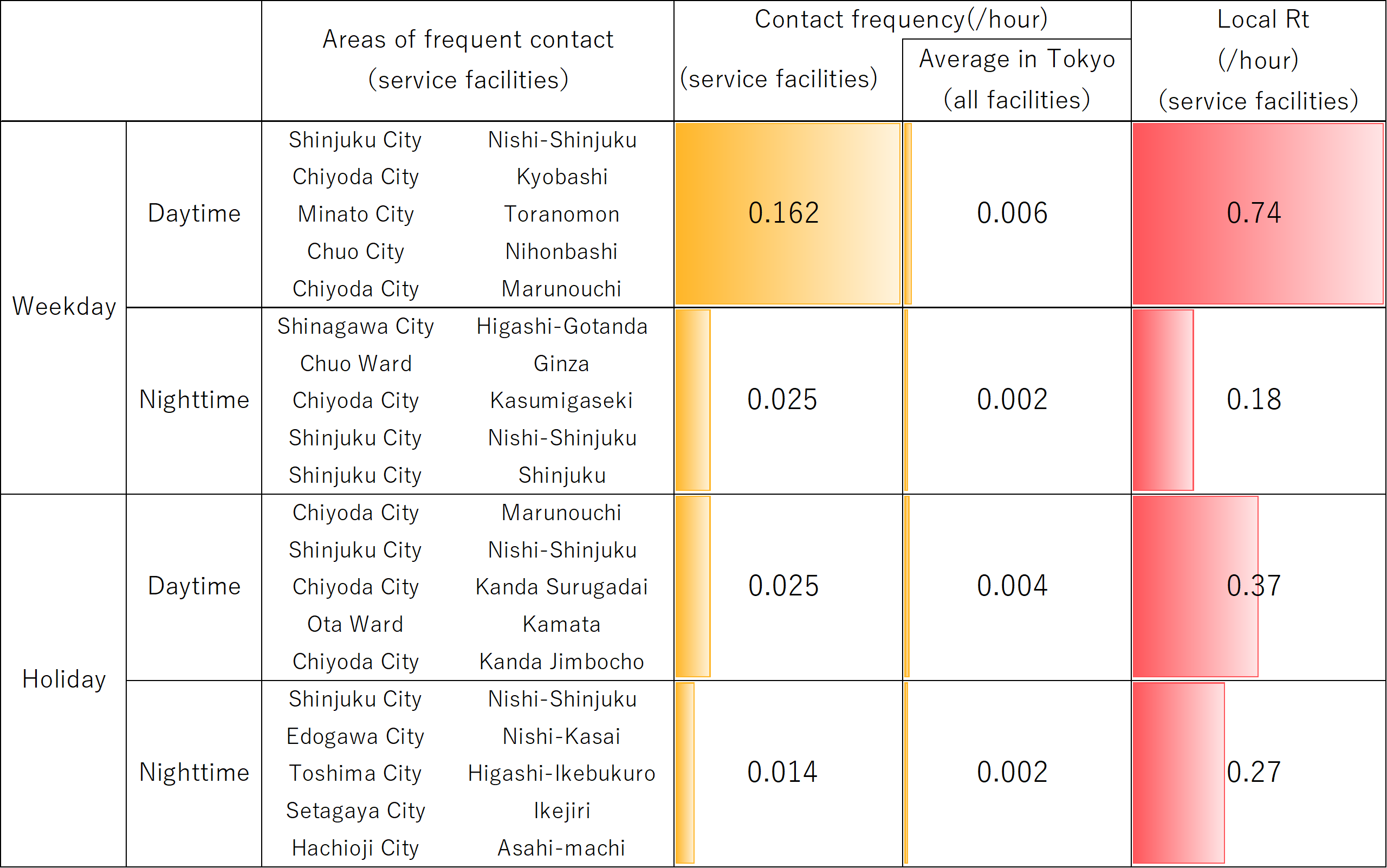}
\caption{
Frequency of people's contacts by weekday/holiday and daytime/nighttime on the service facilities at areas where people are in frequent contact. Average value of contact frequency in Tokyo and the local reproduction number $R_t$ are also shown. 
}
\label{fig:stream}
\end{figure}

\begin{figure}[ht]
\centering
\includegraphics[width=\linewidth]{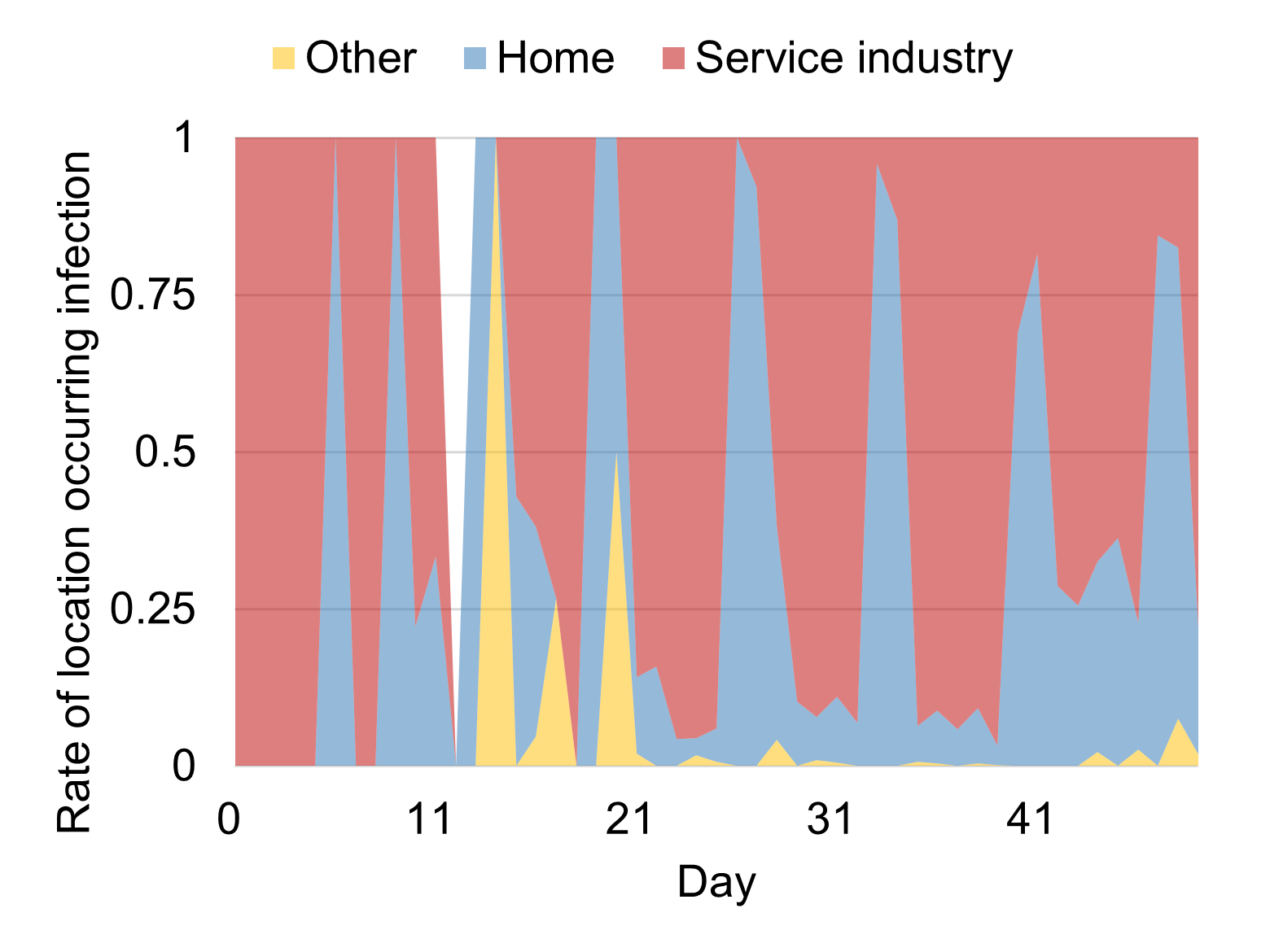}
\caption{
Time dependence of the share of each infection location in the total number of infections obtained from the calculations.
}
\label{fig:stream}
\end{figure}

\begin{figure}[ht]
\centering
\includegraphics[width=\linewidth]{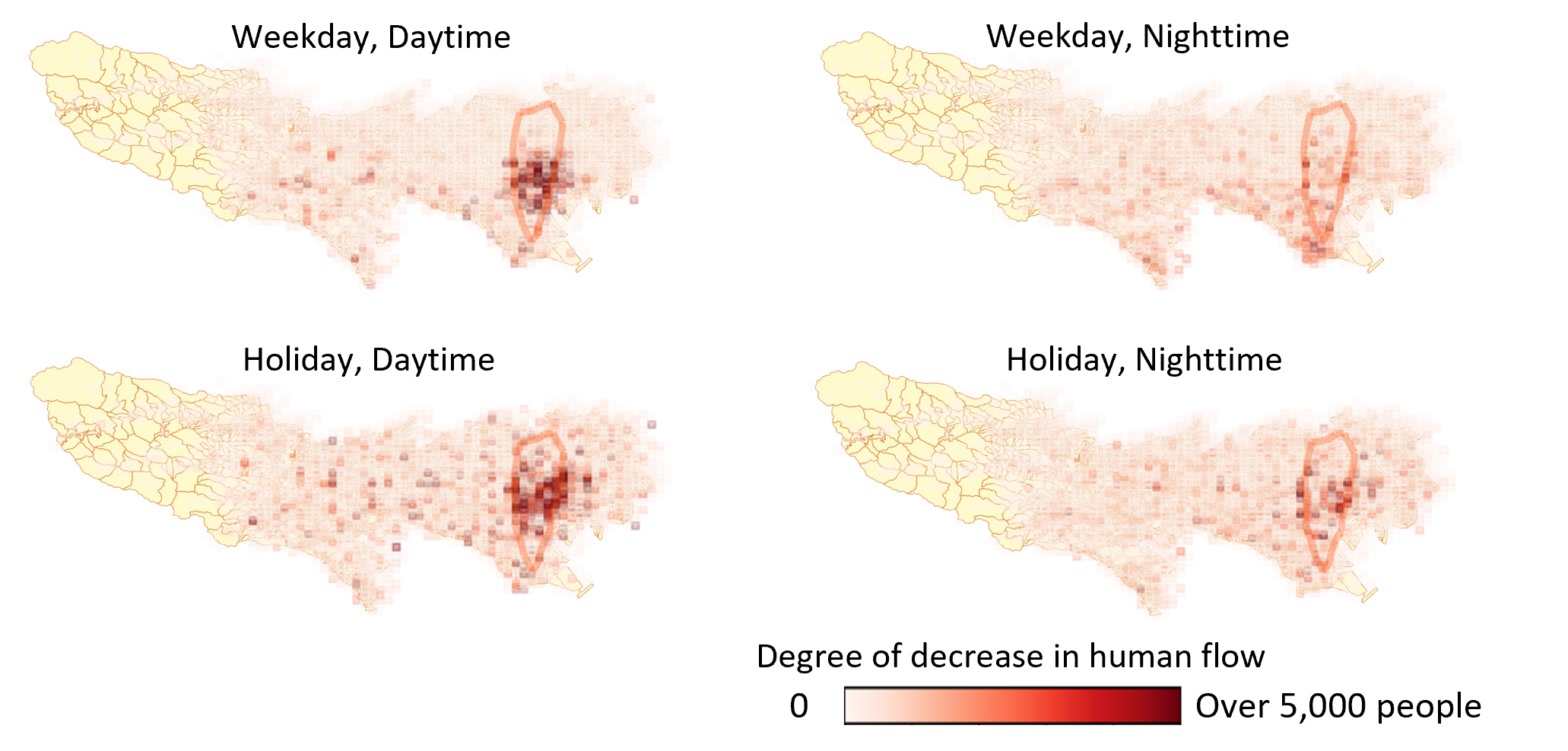}
\caption{
Comparison of Tokyo's human flow in the first wave and a year before, with a heat map showing the degree of decline that occurred after SARS-CoV2. The circles in the map indicate the Yamanote Line (Tokyo's loop train line), i.e., the downtown area.
}
\label{fig:stream}
\end{figure}

\begin{figure}[ht]
\centering
\includegraphics[width=\linewidth]{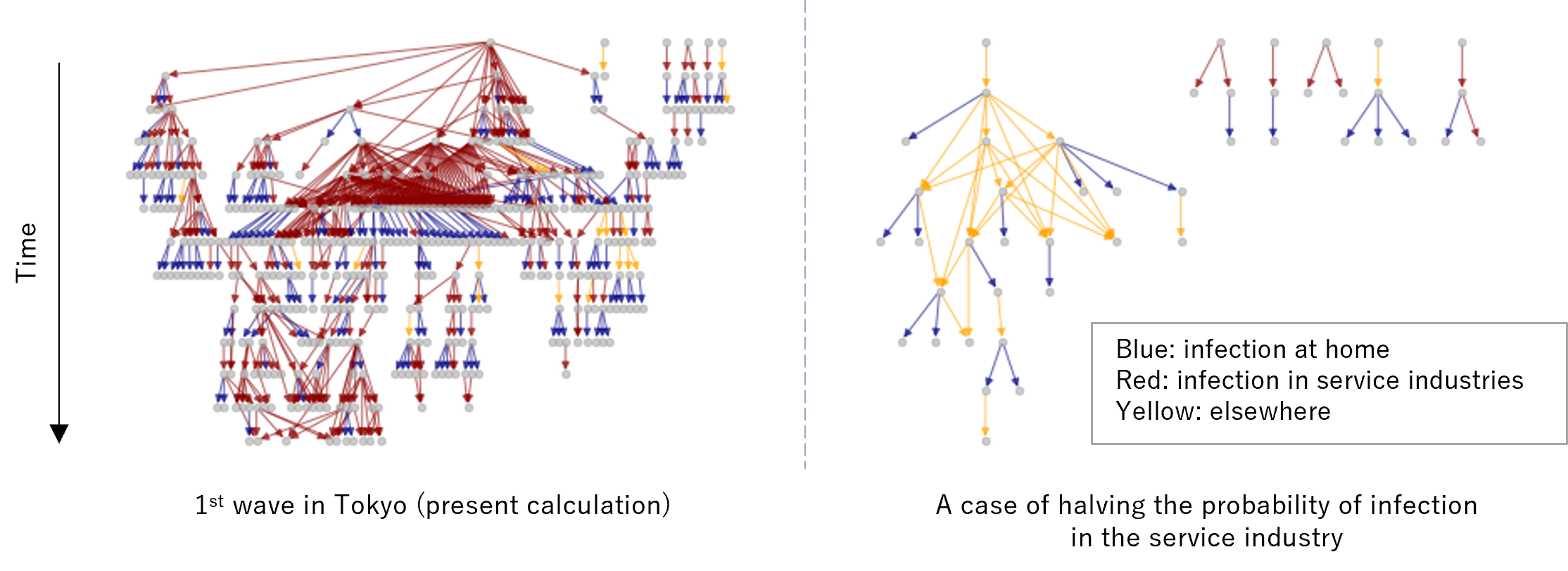}
\caption{
Infection network for agents. The left figure shows the infection network in Figure 1. The right figure shows the infection network in the calculation results assuming that infection countermeasures have been taken in the service industry and the infection risk has been reduced.
}
\label{fig:stream}
\end{figure}

\begin{figure}[ht]
\centering
\includegraphics[width=\linewidth]{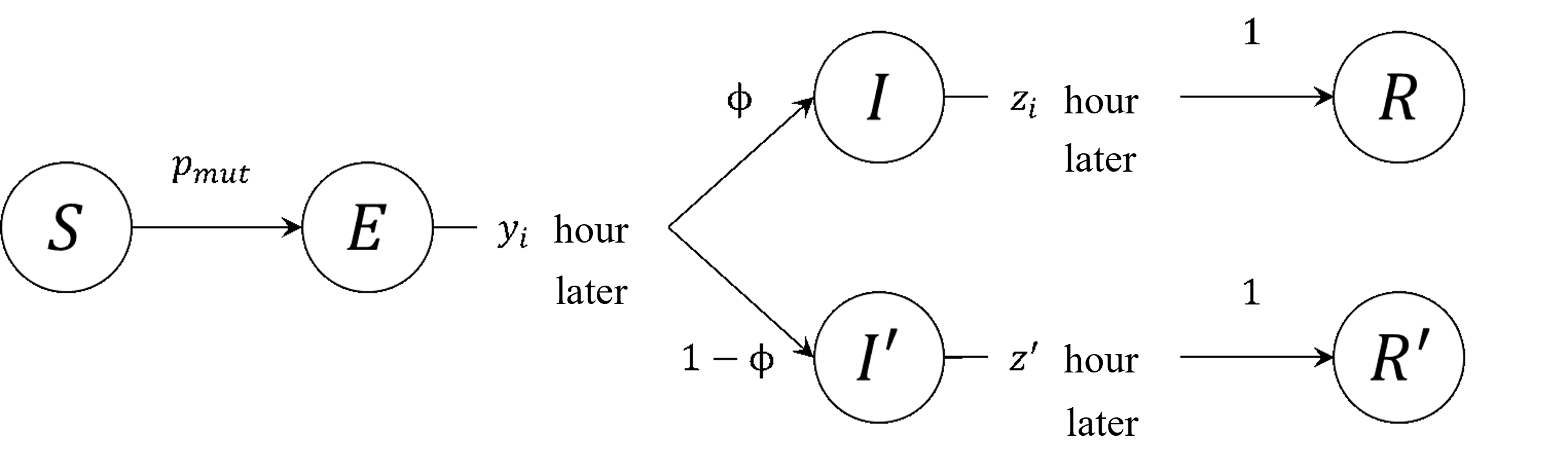}
\caption{
Agent infection status.
S, E, I, and R state denote the infection states in the general SEIR model, respectively. I and R state denote asymptomatic infected agents, while I' and R' denote symptomatic infected agents. I and I' are determined to transition from E state by the probability $\phi$.
}
\label{fig:stream}
\end{figure}

\end{document}